\documentclass[12pt]{article}
\title{ The Effect of On- and Off-Ramps Positions on The Traffic Flow Behaviour} 
  
\author{ Abdelaziz Mhirech$^a$$^,$$^*$, Hamid Ez-Zahraouy$^a$ and Assia Alaoui Ismaili$^b$ .}  
\begin{document}
\maketitle
\begin{center}
{\it \small
Universit\'e Mohammed V, Facult\'e des Sciences, B.P. 1014, Rabat, Morocco.\\$^a$D\'epartement de Physique, Laboratoire de Magn\'etisme et de la Physique des Hautes \'energies. .\\ $^b$D\'epartement de Math\'ematiques.}
\end{center}

\abstract{ The effect of the position of on-ramp  and off-ramp $i_1$ and $i_2$, respectively, on the one dimensional-cellular automaton  traffic flow behaviour, is investigated numerically. The injection rates at $i_1$ and $i_2$ are $\alpha_0$ and $\beta_0$, respectively. However, in the open boundary conditions, with injecting  and extracting rates $\alpha$ and $\beta$ and using parallel dynamics, several phases occur; namely, low density phase (LDP), intermediate density phase (IDP), plateau current phase (PCP) and high density phase (HDP). It is found that phase diagrams exhibit different kind of topologies. For intermediate value of extracting rates $\beta_0$ and $\beta$ and low value of $\alpha$,  $(i_1,\alpha_0)$ phase diagram  provides LDP-IDP, LDP-PCP, IDP-PCP,  and PCP-HDP transitions, and critical end points. The off-ramp position is located to the middle of the road. By increasing $\beta_0$ and $\beta$, the IDP desappears. For high value of $\beta$, only LDP-HDP persists.}\\\\
Pacs number :  05.40.-a, 05.50. +q, 64.60.Cn, 82.20.wt
\------------------------------------------\\
$^*$corresponding author e-mail address: mhirech@fsr.ac.ma

\newpage
\section{Introduction}
\setlength{\parskip}{.2in} 
During the last years, the field of transport have attracted several researchers [  ]. This interest is due primarily to the fact that transportation problems are related to the global behaviour of systems with many elements interacting at short distances, such as the vehicles traveling on the streets, or informations which travel over the internet network. In particular,
the investigation of open traffic systems with on- and  off-ramps is quite popular at the moment [  ]. One reason for this is the impact of the understanding of varying the different flow ratesin order to optimize the total flow or trip times.\\

Among the different methods of investigation and simulation of highway traffic, assymetric simple exclusion process (ASEP) is the most promesing [  ]. Indeed, ASEP is the simplest driven diffusive system where particles on a one-dimensional lattice hop with asymetric rates under excluded volume constraints.\\

The question we want to answer is the following: let us suppose we have a highway running in an urban conglomeration and that there are an access from urban conglomeration to the highway and an exit from the highway. We want to understand where the access and the exit positions must be located in order to maximize the flux of cars in the road. Our aim in this paper is to study the effect of the on-ramp  and off-ramp positions on the one dimensional-cellular automaton  traffic flow behaviour in the open boundaries case. Depending on the injecting and extracting rate values, an adequate localization of the on- and off-ramp positions leads to the appearance of new phases and topologies. Moreover, to compare our results to those where only one off-ramp was taken into account [30], quantitative differences can be understood from the behaviour of average density, current and phase diagrams for different parameters.\\

The paper is organised as follows: Model and method are given in section 2; section 3 is reserved to results and discussion; the conclusion is presented in section 4.

\section{Model}We consider a one-dimensional lattice of length L. Each lattice site is either empty or occupied by one particle. Hence the state of the system is defined by a set of occupation numbers  $\tau_{1}$,$\tau_{2}$,...,$\tau_{L}$, while $\tau_{i} = 1$ ($\tau_{i} = 0$) means that the site i is occupied (empty). We suppose that the main road is single lane, an on-ramp and an off-ramp connect the main road only on single lattice $i_{1}$ for entry and on single lattice $i_{2}$ for way out. During each time interval $\Delta t$, each particle jump to the empty adjacent site on its right and does not move otherwise ($i\neq i_{2}$). $\Delta t$ is an interesting parameter that enables the possibility to interpolate between the cases of fully parallel ($\Delta t =1$) and random sequential ($\Delta t \rightarrow 0$) updates [29]. Particles are injected, by a rate $\alpha \Delta t$, in the first site being to the left side of the road if this site is empty, and particles enter in the road by site $i_{1}$, with a probability $\alpha_{0} \Delta t$ without constraint, if this site is empty. While, the particle being in the last site on the right can leave the road with a rate $\beta \Delta t$ and particles removed on the way out with a rate $\beta_{o} \Delta t$. At site $i_{1}$ ($i_{2}$) the occupation (absorption) priority is given to  the particle which enter in the road (particle leaving the road). Hence the cars, which are added to  the road, avoid any collision.\\
 In our numerical calculations, the rule described above is updated in parallel, $\Delta t=1$, i.e. during one update step the new particle position do not influence the rest and only the previous positions have to be taken into account. During each of the time steps, each particle moves one site unless the adjacent site on its right is occupied by another particle. The advantage of parallel update, with respect to sublattice or sequential update is that all sites are equivalent, which should be the case in realistic model with translational invariance.\\
In order to compute the average of any parameter $w$  $(<w>)$, the values of $w(t)$ obtained from $5\times 10^{4}$ to $10^{5}$ time steps are averaged. Starting the simulations from random configurations, the system reaches a stationary state after a sufficiently large number of time steps. In all our simulations, we averaged over $60-100$  initial configurations. For the update step, we consider two sub steps as shown in figure 1: \\ In the first sub step, the sites $i_{1}$ and $i_{2}$ are updated and in the second half, the chain updates. Thus if the system has the configuration $\tau_{1}(t)$, $\tau_{2}(t)$,...,$\tau_{L}(t)$ at time t it will change at time $ t + \Delta t $ to the following:\\ For $i=i_{1}$ , \begin{equation} \tau_{i}(t+ {\Delta t}/{2}) = 1 
\end{equation} 
with probability 
\begin{equation} q_{i}=\tau_{i}(t)+[\alpha_{0}(1-\tau_{i}(t))-\tau_{i}(t)(1-\tau_{i+1}(t))]\Delta t \end{equation}    
and 
\begin{equation}  
\tau_{i}(t+ \Delta t/2) = 0 
 \end{equation}  with probability $1-q_{i}$. 
Where $i_{1}$ and $\alpha_{0}$ denote the position of the entry site and the injection rate, respectively. \\  
For $i=i_{2}$ , 
\begin{equation}   \tau_{i}(t+ \Delta t/2) = 1 
\end{equation}  with probability  \begin{equation} q_{i}=\tau_{i}(t)+[\tau_{i-1}(1-\tau_{i}) -\beta_{0}\tau_{i}(t)]\Delta t \end{equation}
and 
\begin{equation}
  \tau_{i}(t+ \Delta t/2) = 0  
\end{equation}
with probability $1-q_{i}$. 
Where $i_{2}$ and $\beta_{0}$ denote the position of the absorbing site and the absorbing rate, respectively.\\
 For 1$<$i$<$L with $i\neq i_{1}$ and $i\neq i_{2}$,
 \begin{equation} 
  \tau_{i}(t+ \Delta t) = 1 
 \end{equation}
  with probability 
 \begin{eqnarray} q_{i}=\tau_{i}(t)+[\tau_{i-1}(t)(1-\tau_{i}(t))-\tau_{i}(t)(1-\tau_{i+1}(t))]\Delta t  \end{eqnarray}  
and 
 \begin{equation} 
 \tau_{i}(t+ \Delta t) = 0 
 \end{equation}
  with probability $1-q_{i}$. \\\\  
For $i=1$,
 \begin{equation}
  \tau_{1}(t+ \Delta t) = 1
  \end{equation} 
 with probability 
 \begin{eqnarray}
  q_{1}=\tau_{1}(t)+[\alpha(1-\tau_{1}(t))-\tau_{1}(t)(1-\tau_{2}(t))]\Delta t 
 \end{eqnarray} 
 and 
\begin{equation}  
\tau_{1}(t+ \Delta t) = 0 
 \end{equation} 
 with probability $1-q_{1}$. \\   
For $i=L$, 
 \begin{equation} 
 \tau_{L}(t+ \Delta t) = 1
  \end{equation} 
 with probability 
 \begin{eqnarray} 
 q_{L}=\tau_{L}(t)+[\tau_{L-1}(t)(1-\tau_{L}(t))-\beta\tau_{L}(t)]\Delta t,  
\end{eqnarray}
  and
  \begin{equation}  
\tau_{L}(t+ \Delta t) = 0 
 \end{equation}  
with probability $1-q_{L}$.

\section{Results and Discussion}As we have mentioned previously, our aim in this paper is to study the effect of the positions of   the on- and  off-ramps $i_1$ and $i_2$, respectively, for different values of $\alpha$, $\beta_{0}$ and $\beta$, on the average density and flux in chain. The study is made in the open boundary   conditions case. $\alpha_{0}$ and  $\alpha$ denote the injecting rates at first site (i = 1) and at site $i_{1}$, respectively.  $\beta_{0}$ and $\beta$ are the extracting rates at site $i_{2}$ and at the last one (i = L), respectively. The length of the road studied here is L=1000. \\ 
The figures 2(a) and 2(b) give respectively the variation of the average density $\rho$  and average  current $J$  versus the injecting rate $\alpha_0$ for several values of the on-ramp position $i_1$. These figures are given for  $\alpha = 0.1$, $\beta = 0.1$ and $\beta_0 = 0.4$. However, when the on-ramp is located upstream of the off-ramp, the system studied exhibits four phases, depending of the behaviours of the density, $\rho$, and the current, $J$. Namely: i) The low density phase (LDP), where the averages density and  current increase when increasing the rate of injected particles $\alpha_{0}$. ii) The intermediate density phase (IDP) characterised by a smoothly increase of the density and average current. iii) The plateau  current phase (PCP) for which  the density and current are constant in a special interval of $\alpha_0$. iv) The high density phase (HDP) in which, for high values of $\alpha_0$, the current decreases and  the density reaches its maximum value and remains constant. On the other hand, when $i_1$ is located downstream from $i_2$, the IDP desappears.  In addition, when increansing $i_1$ for a given value of $\alpha_0$, the figure 2(a) shows that the average density is constant in LDP, increases in IDP and PCP then decreases in HDP. The figure 2(a) exhibits an inversion point situated at the PCP-HDP transition. Moreover, the figure 2(b) shows that the average current deceases by increasing $i_1$, for any value of $\alpha_0$.\\
 For $i_1 < i_2$, we note that the IDP, which doesn't appear in the model where only the off-ramp is taken into account [30], occurs for the intermediate values of  $\alpha_0$ ($\alpha_{0c1}<\alpha_0<\alpha_{0c2}$). $\alpha_{0c1}$ and $\alpha_{0c2}$ correspond to the transition between LDP-IDP and IDP-PCP, respectively. While the PCP arises between two critical values $\alpha_{0c2}$ and $\alpha_{0c3}$ of injecting rate $\alpha_{0}$. Where  $\alpha_{0c3}$ corresponds to the PCP-HDP transition. Note that the  transitions which occur at $\alpha_{0c3}$  disappears when $i_1$ is located after $i_2$ (Figure 2a). Now, in order to have a suitable criterion for determination of the nature of the transition, we identify the first order transition (abrupt transition) by the jump in the average density or by the existence of a peak in the derivative of $\rho (\alpha_{0})$ with respect to $\alpha_{0}$. The jump in density corresponds to a first order transition [29]. This means that the  above transitions are of first order type.\\
Collecting the results illustrated in figures 2(a) and 2(b), the four regions are given on the phase diagram ($i_1$,$\alpha_{0}$) shown in Figure 2(c). Beside this, such phase diagram  exhibits four critial end points, around which there is no distnction between the phases. This critical end points are indicated by $CEP$.\\  
For low values of $\alpha$, $\beta$ and $\beta_0$ ($\alpha=0.1$, $\beta=0.1$ and $\beta_0=0.1$), The ($i_1,\alpha_0$) phase diagram is presented in figure 3. This figure exhibits only tow first order phase transitions. Namely, LDP-PCP, PCP-HDP transitions. The later one can be founed by variying $\alpha_0$,for a given value of $i_1$ lower than $i_2$, or at $i_1=i_2$, for $0.45<\alpha_0<0.95$. Moreover, the figure 3 exhibits two critical end points. The comparison of figures 2 and 3 highlight the effect of $\beta_0$ on the ($i_1,\alpha_0$) phase diagram. Indeed, for intermediate value of $\beta_0$, the IDP arises.\\
For a sufficiently large value of $\beta_0$, the critical end points desapears, as shown in figure 4. This figure is given for $\alpha=0.1$, $\beta=0.1$ and $\beta_0=0.8$.\\
The figure 5 give the ($i_1,\alpha_0$) phase diagram for $\alpha=0.1$, $\beta_0=0.1$ and $\beta=0.3$. In this case, the system exhibits four phases and three critical end points. From figures 2c and 5, we deduce that the IDP arises for intemediate values of extracting rates $\beta$ or $\beta_0$, when $i_1$ is upstream of $_2$.\\

 \section{Conclusion}  Using numerical simulations, we have studied the effect of the  on- and off-ramp positions on the traffic flow behaviour of a one dimensional-cellular automaton, with parallel update. Depending on the values of $\alpha$, $\beta$ and $\beta_0$, the ($i_1,\alpha_0$) phase diagram exhibits different topologies. The IDP occurs only at special positions of on- and off-ramps with special values of extracting rates $\beta$ and $\beta_0$ and injecting rates $\alpha$ and $\alpha_0$. The transition between different phases are of first order. Furthermore, the system exhibits critical end points in the ($i_1,\alpha_0$) plane in the case of moderate values of $\beta$ and $\beta_0$ and small value of $\alpha$.\\

\newpage
\textit{Figure captions:}

 Fig.1: Example of configuration obtained after two sub steps for system size $L=15$. \\
 Fig.2: For $\alpha=0.1$ and $\beta=0.1$; (a) Average density $\rho$ versus the injection rate $\alpha_{0}$;
 (b) Variation of the average current as a function of $\alpha_{0}$; (c) Phase diagram
 ($\beta_{0}$,$\alpha_{0}$). The number accompanying each curve, in (a) and (b), denotes the values of $\beta_{0}$.\\ 
 Fig.3: Phase diagram ($\beta$,$\alpha_{0}$) for $\alpha=0.1$ and $\beta_{0}=0.1$.\\
 Fig.4: Phase diagrams ($\beta_{0}$,$\alpha_{0}$) for $\beta=0.1$; (a) $\alpha=0.2$; (b) $\alpha=0.4$; (c) $\alpha=0.5$.\end{document}